\documentclass[prb,twocolumn]{revtex4}

\usepackage{amsmath}
\usepackage{graphicx}

\begin{document}

\title{A new equivalence between fused RSOS and loop models}

\author{Lukasz Fidkowski$^{1,2}$}
\affiliation{$^1$Microsoft Station Q, University of California, Santa Barbara 93106-6105\\
$^2$ Department of Physics, Stanford University, Stanford, CA 94305}

\date{\today }

\begin{abstract}
We consider the topological theories of [\onlinecite{LWstrnet}] and [\onlinecite{FFNWW}] and study ground state amplitudes of string net configurations which consist of large chunks $G$ of (trivalent) regular lattice.  We evaluate these amplitudes in two different ways: first we use the Turaev-Viro prescription to write the amplitude as a sum over labelings of the faces of $G$, and second we use the local rules that constrain the amplitude (the $F$-matrix) to resolve subgraphs in creative ways.  In the case of the Doubled Fibonacci theory this second way allows us to produce loop models.  In particular, we show that the hard hexagon model is equivalent to an anisotropic loop model.  Many other interesting equivalences can presumably be obtained.
\end{abstract}

\maketitle
\section{Introduction}

\begin{figure}[tbh!]
\includegraphics[width=2.2in]{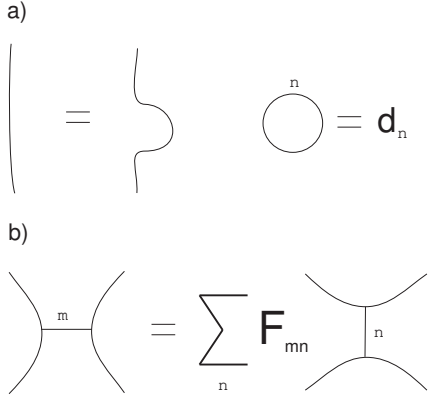}
\caption{Local relations for ground state amplitudes.  a) $d$-isotopy invariance.  Here $d_n$ is an invariant associated with each particle type $n$ and called its quantum dimension  b) $F$-matrix relations.  In both a) and b), by each graph we implicitly mean the ground state amplitude of that graph}
\label{fig1}
\end{figure}

In [\onlinecite{LWstrnet}] and [\onlinecite{FFNWW}] two dimensional models realizing nonabelian topological phases of matter are introduced.  These models are formulated in terms of string nets, which are trivalent graphs labeled by the particle types $0, \ldots, N-1$ of the theory.  States in Hilbert space are specified by assigning an amplitude to each string net configuration.  The ground state amplitudes of these models turn out to satisfy certain local rules - specifically $d$-isotopy and the $F$-matrix constraints (see fig. \ref{fig1}).

We will call the ground state amplitude assigned to a net $G$ its quantum evaluation and denote it $\langle G \rangle$.  One can show that on a plane (or equivalently a sphere) $\langle G \rangle$ is uniquely determined once we stipulate $\langle \text{empty graph} \rangle = 1$.  This is done by repeatedly using $F$-matrix and $d$-isotopy moves to reduce a graph down to the empty net (this is not true on a surface of nontrivial topology, where one has multiple ground states).  

\section{Turaev-Viro evaluation}

From now on we will be working with planar nets that are located in a bounded region of the plane.  Thinking of these as graphs on the sphere, we call the region of $G$ (i.e. face of $G$, or vertex of the dual graph $\hat{G}$) that contains the point at infinity the region at infinity.  One way to evaluate $\langle G \rangle$ makes use of the fact that it is just the expectation value of ``Wilson nets" in the Chern-Simons theory describing the infrared limit of our model.  The work of Turaev and Viro, also called the ``shadow method" \cite{Kauffman}, shows that:

\begin{equation}
\label{eqn1}
{\langle G \rangle} = \sum_{L_f} w(L_f)
\end{equation}

\begin{figure}[tbh!]
\includegraphics[width=3.2in]{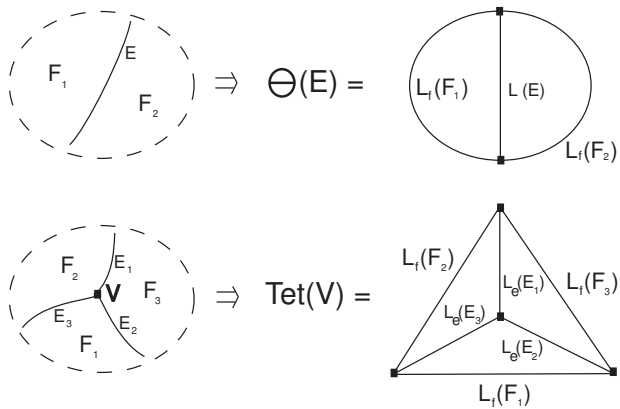}
\caption{Definition of the theta and tetrahedral symbols}
\label{fig2}
\end{figure}

Let us explain the meaning of this formula.  It will be necessary to first introduce some more notation.  The sum in (\ref{eqn1}) is over the labelings $L_f$ of the faces of $G$ by the particle types of $0, \ldots, N-1$ of the theory, with the caveat that we fix the label of the region at infinity to be $0$ (the trivial particle type).  The weight $w(L_f)$ is defined as follows.  Let $L_f(F)$ be the particle type that labels face $F$.  Recall that we also have a labeling of the edges, which we call $L_e(E)$.  Now, for each edge $E$ we can define $\Theta(E)$ and for each vertex $V$ we can define the tetrahedral symbol $\text{Tet} (V)$.  These are given as the quantum evaluations of the appropriate theta and tetrahedral graphs involving the edges and faces adjacent to the appropriate vertex or edge (see fig. \ref{fig2}).

Notice that the tetrahedral and theta symbols are not always well defined - there is a condition on which particle types are allowed to come together at a trivalent vertex.  If all tetrahedral and theta symbols are well defined, the labeling is called admissible; quick inspection of fig. \ref{fig2} shows that it is sufficient to check that all the theta graphs are well defined.  We note that this is precisely the same situation as in a fused RSOS model whose degrees of freedom live on the faces of $G$ \cite{F}.

The weight of an inadmissible labeling is defined to be $0$; that of an admissible one is:

\begin{equation}
\label{eqn2}
w(L_f) = \prod_{\text{faces } F} d_{L_f (F)} \prod_{\text{edges } E} {\Theta(E)}^{-1} \prod_{\text{vertices } V} {\text{Tet} (V)}
\end{equation}

We have thus expressed the quantum evaluation as the partition function of an RSOS-like model with local Boltzmann weights.  Before we work out the implications and specific examples of this representation, we motivate further the origin of (\ref{eqn1}).

Take the $3$ dimensional ball $B$ with boundary $S^2$ and an associated topological quantum field theory.  We can use the Turaev-Viro formalism to evaluate the expectation value of any (labeled) string net $G$ on $S^2$ as follows (for simplicity, we describe the procedure for unoriented theories).  First, we pick a cellulation of $B$ compatible with $G$ - that is we subdivide $B$ into cells.  Compatibility with $G$ is just the condition that the intersection of the boundary $S^2$ with the $2$-skeleton of the cellulation is $G$.  For each labeling $L$ of the $2$-skeleton compatible with the labeling of the edges of $G$ (this means that we label the $2$-cells which contain an edge of $G$ with the label on that edge) we define a weight $w(L)$ in a fashion analogous to (\ref{eqn2}): each $2$-cell contributes a factor of a quantum dimension, each $1$-cell an inverse theta symbol, and each $0$-cell a tetrahedral symbol.  The Turaev-Viro prescription gives the formula for the expectation value of $G$ as a sum over admissible labelings, with the caveat that we hold one pre-chosen region fixed with the label $0$ (this turns out to be just a normalization condition).

Equation (\ref{eqn1}) follows as a special case of this formula.  We obtain it by taking a specific cellulation of $B$.  The cellulation we choose is as follows: let $M$ be a smaller sphere, concentric with $\partial B = S^2$ and with radius $r < 1$.  The string net $G$ lives on $\partial B$.  By ``dropping" $G$ down to $M$ we sweep out a $2$-dimensional surface (with singularities); taking its union with $M$ we obtain the $2$-skeleton of our cellulation.  We then see that the $2$-cells in this cellulation correspond to either edges of $G$ or faces of $G$, and (\ref{eqn1}) follows from (\ref{eqn2}).

\section{Loop Models}

\begin{figure}[tbh!]
\includegraphics[width=3in]{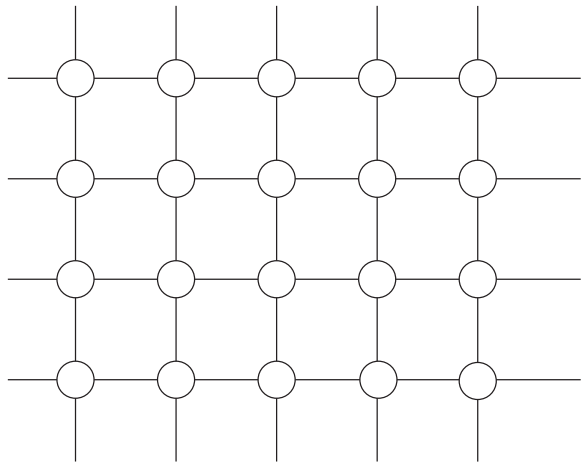}
\caption{A distinctive trivalent lattice}
\label{fig6}
\end{figure}

To illustrate the second way of computing the quantum evaluation we specialize to the case of the doubled Fibonacci theory, which has two particle types: $0$ (trivial) and $1$ (nontrivial).  We can stick to just drawing nontrivial edges, and hence reduce to considering only unlabeled string nets - the restriction on the particle types allowed to fuse at a vertex just becomes the constraint of having no univalent vertices.  The essence of the second approach is to use the $F$-matrix and $d$ isotopy relations to resolve all the trivalent vertices in a regular lattice.  We start with a string net that looks like a large chunk of the configuration in fig. \ref{fig6}.

\begin{figure}[tbh!]
\includegraphics[width=2.5in]{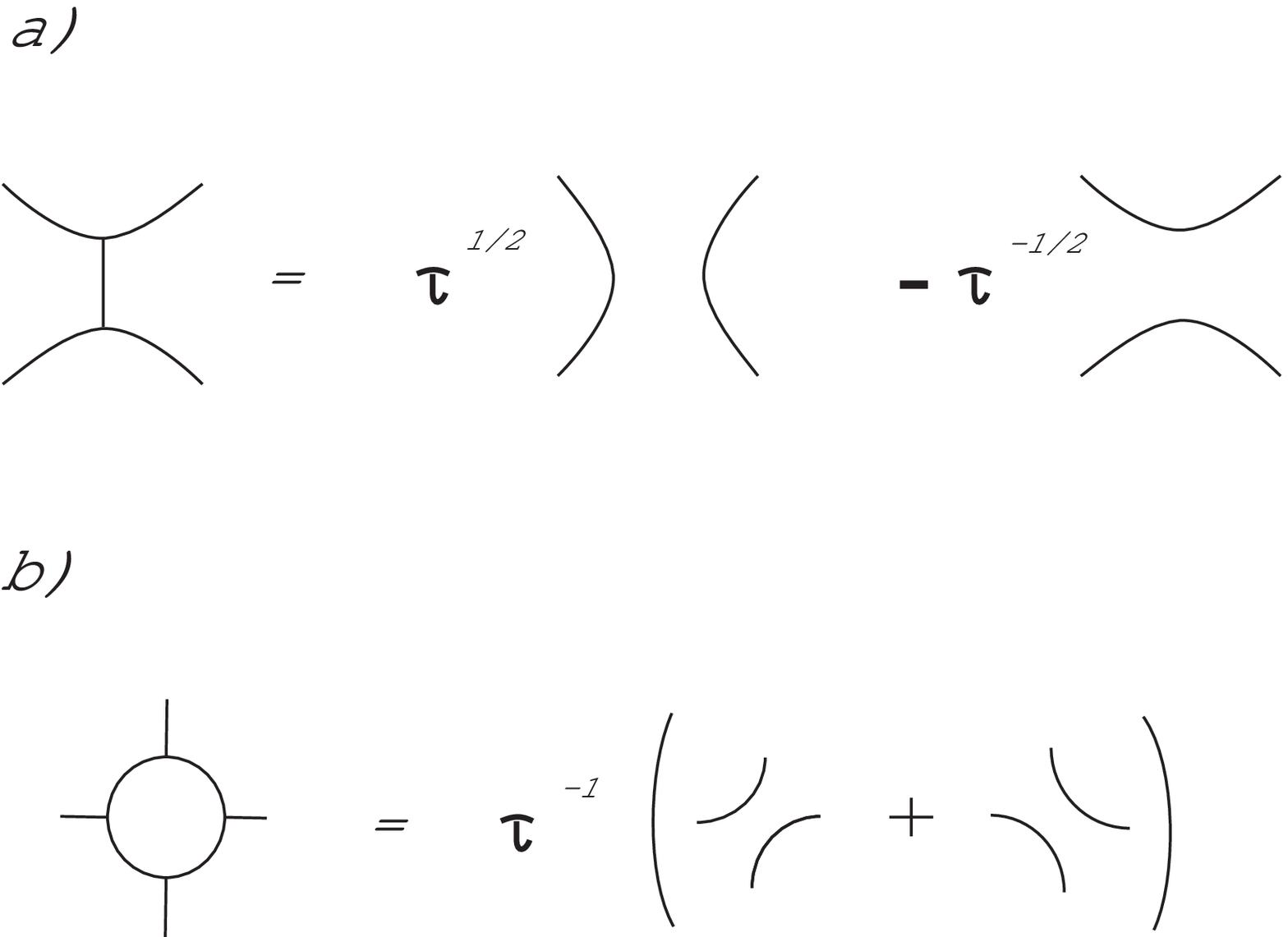}
\caption{Local identities that are useful for resolving trivalent vertices}
\label{fig4}
\end{figure}

We now resolve the trivalent vertices using the identity in fig. \ref{fig4} b) (note: $\tau = (1 + \sqrt{5})/2$).

\begin{figure}[tbh!]
\includegraphics[width=3in]{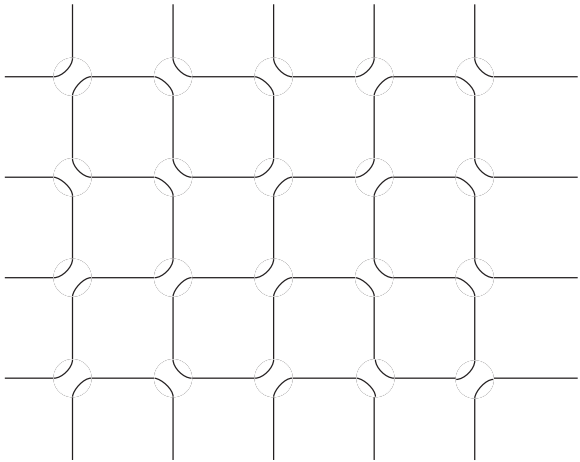}
\caption{The quantum evaluation is expressed as a sum over resolutions such as this one}
\label{fig7}
\end{figure}

We thus have two possibilities for resolving each circle, and get an expression for the quantum evaluation as a sum over fully packed loops, with a factor of $\tau$ for each loop.  One such sample configuration is shown in fig. \ref{fig7}.  The factor of $\tau^{-1}$ out front contributes  only to an overall vertex fugacity.

On the other hand, the Turaev-Viro prescription gives the quantum evaluation as a sum over labelings of the faces of the lattice with $0$'s and $1$'s.  In order for a labeling to be admissible we just can't have two adjacent regions labeled with a $0$.  It is useful to think of the regions labeled with a $0$ as being occupied by a particle, so that the constraint is that we're not allowed to have particles on adjacent regions.  The local Boltzmann weight turns out to simply be determined by a particle fugacity (the fugacities of the octagons and circular regions are different), with an overall normalization constant whose logarithm scales with the number of regions.  This is all analyzed in detail for a different lattice in Appendix A of [\onlinecite{FFNWW}].

We have thus shown that two different models - a fused RSOS model and a fully packed loop model - have the same partition function.  Presumably we can extend this to a correspondence between the operators of the models and show that the two are actually equivalent.  Equivalences of this type have been recently studied \cite{F}, and our result seems to be along similar lines as that of [\onlinecite{F}].  However, we haven't carefully studied the precise connection.

\begin{figure}[tbh!]
\includegraphics[width=2.5in]{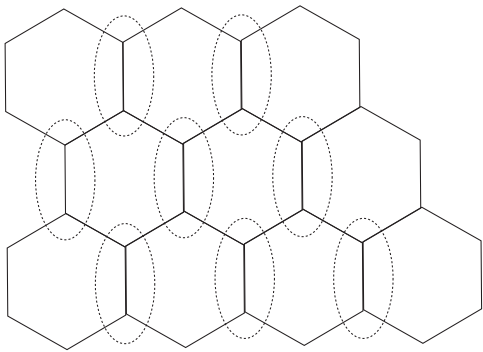}
\caption{The hexagonal lattice.  The circled portions are to be resolved}
\label{fig3}
\end{figure}

\begin{figure}[tbh!]
\includegraphics[width=2.5in]{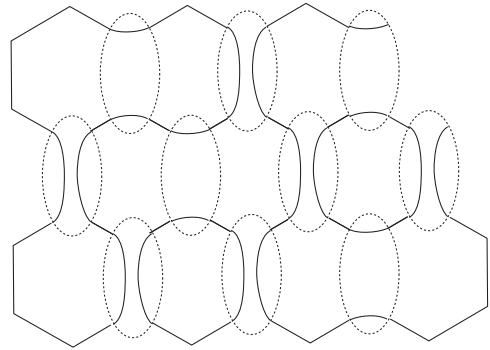}
\caption{Sample loop configuration from resolving the hexagonal lattice}
\label{fig5}
\end{figure}

Another example comes from considering the hexagonal lattice (fig. \ref{fig3}).  We use the identity in fig. \ref{fig4} a) to resolve the circled portions of the lattice, and obtain the amplitude as a sum over loop configurations such as that in fig. \ref{fig5}.  This time the weighting of the loops isn't quite topological because of the coefficients in fig. \ref{fig4} a).  However, all the weights are positive, because each minus sign comes in an even number of times.  We obtain an anisotropic model of loops.

The Turaev-Viro prescription, on the other hand, shows that the corresponding RSOS model is just the critical hard hexagon model.  We believe this is general and all models obtained in this way will be critical.  So we have an equivalence between the critical hard hexagon model and an anisotropic model of loops.  Note also that we can use the identity in fig. \ref{fig4} a) to resolve all the trivalent vertices of any graph that admits a dimer covering.

\section{Conclusions}
We found an equivalence between some fused RSOS models and loop models.  We believe the models involved will always be critical \cite{EW}.  This equivalence is probably closely related to that of [\onlinecite{F}].  In principle we can be more general: in a general topological theory with a given lattice string net, we can be find creative ways of applying local relations and obtain nontrivial models.  It would be interesting to explore this equivalence further, and also extend it past simple equality of partition functions.  In particular, one would like to study the conformal field theories describing the critical points.

\section{Acknowledgements}
I would like to thank Paul Fendley and Michael Freedman for useful discussions and especially Kevin Walker for originally suggesting the shadow method as a way of computing quantum evaluations.  This research was supported partly by the NSF under grant no. PHY $-0244728$.

\end{document}